\documentstyle[12pt]{article}
\def\eq{\begin{equation}}
\def\en{\end{equation}}
\newcommand \be  {\begin{equation}}
\newcommand \bea {\begin{eqnarray} \nonumber }
\newcommand \ee  {\end{equation}}
\newcommand \eea {\end{eqnarray}}
\def\ceq{C_{eq}}
\def\req{r_{eq}}
\def\cas{C_{as}}
\def\ras{r_{as}}
\def\bas{b_{as}}
\def\mas{m_{as}}
\def\Mb{\overline{M}}
\def\ol{\overline}

 \def\(({\left(}
 \def\)){\right)}
\def\bi{\bibitem}
\def \la{\langle}
\def\ra{\rangle}
\def \a{\alpha}
\def\jpa{J.Phys.A}
\def\d{{\rm d}}
\def\beqna{\begin{eqnarray} \nonumber}
\def\eeqna{\end{eqnarray}}
\def\ov{\over}
\def\l{\left}
\def\r{\right}
\def\tq{\tilde{q}}

\def \cC{{\cal C}}
\def \mag{{\cal M}}
\def\bphi{{\mbox{\boldmath{$\phi$}}}}
\def\bpsi{{\mbox{\boldmath{$\psi$}}}}
\def\a{\alpha}
\def\b{\beta}
\def\nn{\nonumber}
\def\de{\delta}

\newcommand{\pub}[4]{{\em #1 }{\bf #2}, #3 (#4)}

\begin{document}
\topmargin -1.0 true cm
\textheight 23 true cm
\baselineskip .5 true cm
\begin{titlepage}

\vskip .27in
\begin{center}

{\large \bf On mean field glassy dynamics out of equilibrium}
\vskip .27in

Silvio Franz and Marc M\'ezard
\vskip .2in

{\it  Laboratoire de Physique Th\'eorique de l'Ecole
Normale Sup\'{e}rieure \footnote {Unit\'e propre du CNRS,  associ\'ee
 \`a\ l'Ecole
 Normale Sup\'erieure et \`a\ l'Universit\'e de Paris Sud, 24 rue
 Lhomond, 75231 Paris Cedex 05, France}}

\end{center}
\vskip 8pt

{\bf Abstract }
We study the off equilibrium dynamics of a mean field disordered systems
which can be interpreted both as a long range interaction spin glass and
as a particle in a random potential. The statics of this problem is
well known and exhibits a low temperature spin glass phase
with continuous replica symmetry breaking.  We
study the equations of off equilibrium dynamics with analytical
and numerical methods.  In the spin glass
phase, we find that the usual equilibrium dynamics (observed when
the observation time is much smaller than the waiting time) coexists
with an aging regime. In this aging regime, we propose a solution implying
 a hierarchy of crossovers between the observation time
and the waiting time.

\vfill
{ {\bf \footnotesize
LPTENS preprint 94/05}}

{ {\bf \footnotesize
cond-mat/9403004}}
\vfill
\end{titlepage}

\newpage
\section{Introduction}
A lot of efforts have been
devoted  in the last fifteen years
to the study of equilibrium static and dynamic properties
of spin glasses \cite{mpv,binyou,fishertz}.
Comparatively, the off equilibrium dynamical effects have received
less attention. The recent years have seen a renewal of interest for this
OED.
One  reason is experimental. While it is clear that many experimental
observations are inherently dynamical effects, the status of the
off equilibrium
dynamical effects have turned recently from that of an annoying
perturbance to that of a very powerful probe. Some
of the most interesting recent experimental
findings in spin glasses, like the slow relaxation of the thermoremanent
magnetization, aging, and memory effects during temperature cycling
experiments, are inherently out of equilibrium phenomena
\cite{AGESWE,AGEUSA,AGEFRA,AGETWO}. Several phenomenological models
of these effects have already been proposed, based on ideas of
droplets \cite{kophil,fishus} or phase space traps
with a  broad distribution of trapping times \cite{bouchaud}.
 The second origin of this upsurge of interest
comes from the theoretical side. Prompted by the experimental observations,
it has been realized recently
 that some  microscopic analytical approach
to these problems is possible, and that the  off equilibrium
nature of the
dynamics might even cure some old problems of the dynamical approach.
The first works on spin glass dynamics, following the idea that the
use of a dynamical generating functional could be an alternative to
the introduction of replicas \cite{dedom}, focused on the
ED \cite{sz,sompo}. Early attempts to model
some aspects of the OED  along these same
lines have concentrated
on the mean field theory of spin glasses close to the critical
temperature, taking into account explicitely the changes
in external parameters like temperature or magnetic field \cite{horner,dfi}.
More recently, it has been
observed that these  effects can be studied
without any reference to time variation of the external parameters, but
by keeping into account  the existence of an initial time
for the dynamics (corresponding to the quench into the spin glass phase
in the experiments), and the existence of a finite waiting time
\cite{opper,cuku,rieger,CUKURI,mapa}.

In this paper we study the off equilibrium dynamics (OED) through a
microscopic approach along the lines above. We consider the
problem of an oriented D dimensional manifold
 embedded in a D+N dimensional space,
in presence of a random potential. This is a very interesting
and general problem  \cite{manifrev} which is connected to
interface pinning
by impurities , directed polymers in disordered
media, vortex pinning in high temperature superconductors \cite{fglv,bmy},
and also, after various mappings, to growth phenomena \cite{kpz}
or turbulence \cite{burgers}.
We shall work in the limit of an infinite dimensional embedding
space ($N \to \infty$). This limit has two major advantages. It allows
for the derivation of exact integrodifferential equations for the
correlation and response functions.
 Also in this limit the static properties
have been studied in details using the replica method, and it has been
shown that a full hierarchical replica symmetry breaking (r.s.b.) is needed
in order to describe the system \cite{mp}.

Our work has two aspects. One is an analytic
 study of the OED equations at
large times, which shows a possible family of solutions related
to the static (r.s.b.) solution. The other one is the numerical solution
of these equations. This numerical solution is in fact limited
to the $D=0$ version of the general random
 manifold problem.
This is nothing but the "toy model" of a  single
particle in N dimensions, submitted to a potential which is the sum of
a quadratic well and a Brownian process
 \cite{villain,engel1,mptoy,engel2,mapa}.  In the large $N$ limit,
this model can be interpreted as a long range spin glass model, and
we shall show that
many interesting aspects of the dynamics are kept by this  toy model,
as is true for the statics \cite{mp,mptoy}. A brief account of our
work has appeared recently \cite{fm1}.

The equilibrium Langevin dynamics (ED) of the manifolds
 in the large $N$ limit
has been worked out by Kinzelbach and Horner  in two interesting
recent papers \cite{kh1,kh2}, following the general strategy
used by Sompolinsky and Zippelius \cite{sz,sompo} in spin glasses.
We shall basically reconsider their approach,
using the OED corresponding to a finite
waiting time, in the spirit of the recent work by Cugliandolo
and Kurchan on the spherical spin glass with p-spin interactions
\cite{cuku}. Technically the difference is that in the off-equilibrium
dynamics
 the dynamical evolution starts at a time $t_0=0$. Therefore
the correlation function $C(t,t')$ and the response function
$r(t,t')$ depend explicitely on both $t$ and $t'$. In the equilibrium
dynamics the time $t_0$ is sent to $-\infty$,
and the correlation and response
become functions of the differences between $t$ and $t'$: $C_{eq}(t-t')$
and $r_{eq}(t-t')$.

As we shall see there are many formal similarities between these
two dynamics, together with formal similarities with the static
r.s.b. solution. However one should keep
in mind that  the physical contents of
these two approaches are actually quite different. In
ED, $\ceq$ and $\req$ satisfy coupled equations which
depend explicitely on an anomaly of the response occuring on infinite time
scales. One must assume the existence of a regularization of these
diverging times by considering for instance a system with a finite
number of degrees of freedom. The "dynamical" equations on diverging
time scales turn out to be identical to the static (r.s.b.) equations of the
replica method. It is important to notice that this "dynamics" on diverging
time scales is not really a dynamical solution (for instance it is invariant
under arbitrary reparametrizations of time). In our opinion this
equilibrium  "dynamics", considered on diverging time scales, rather
gives an "intuitive" and appealing description of the strange
algebra of the replica method \cite{cufrakumez}.

In contrast, in OED, $C(t,t')$ and $r(t,t')$ obey
causal equations which have a
unique solution  (for instance, for $t>t'$,
$\partial C(t,t')/\partial t$ depends
only on $C$ and $r$ evaluated at times smaller than t.)\cite{cuku}.
 One can work directly with
 an infinite system, and there is
no need to introduce diverging time scales. An important point is that
the introduction of a finite waiting time provides a natural
regularization: as we shall see, the roles of the diverging time
scales are then played by some functions (e.g. powers) of the waiting time.

It is not easy to get some analytical information on the correlation
and response in OED. However, as they obey
causal equations, one can solve them numerically in a rather
straightforward way.
Our work is based on a detailed numerical solution of these OED
equations of the toy model. We shall divide our results
into two groups. One which refers to the asymptotic regime ($t-t'$ finite),
the other refers to the non asymptotic regime.

In the asymptotic regime, we shall present hereafter numerical
evidence that: 1) There exists a limiting response function $\ras(\tau)=
lim_{t_w \to \infty} r(t_w+\tau,t_w)$; 2) This function is the same as
that derived in ED \cite{kh1}, with a certain condition of
criticality of the anomalous response coming from diverging time scales;
3) Similar results hold for the correlation.
In particular, the values of $\cas(0)$ and $\cas(\tau \to \infty)$
agree with the results for the statics from the full r.s.b. solution;
4) The energy $E(\tau)$ also converges to its static r.s.b. value
at large $\tau$.

These results on the asymptotic behaviour provide  an independent confirmation
of both the static r.s.b. approach, as well as the usual equilibrium
dynamics on finite time scales. In order to understand the origin
of these results, and simultaneously to study the aging effects,
one needs a careful study of the correlation and response for
finite waiting times.
 Here we shall point out a few effects: 1)  The very fact that one recovers
the static  r.s.b. results in the spin glass phase implies that there must
be aging effects (in the sense that, at an arbitrary large time $t$,
some perturbation of the
system at  times $t'<t$ has a relevant effect, even when
$t-t'$ is very large). These aging effects are also seen in our
numerical studies on the (short) time scales we can achieve.
 2) It is possible to find a family
of approximate solutions of dynamical equations at large times. These solutions
are technically related to the solutions of the dynamics on diverging
time scales found in $\cite{kh2}$, but the role of the "diverging
time scales" is now played by some functions of the waiting times
(like for instance $t_w^u$).

In the next section we introduce the model and write down the dynamical
equations in the large N limit. In sect. 3 we review the static results
obtained with the replica method. Sect. 4 presents an analytic study of
the asymptotic regime, which is compared to the numerical integration of
the equations in sect. 5. Sect. 6 deals with the aging regime.
Some perspectives are summarized in sect.7.

\section{The model}

 The manifold is decribed by a $N$ component field $\phi_\alpha(x)$,
where $\alpha \in {1,...N}$. The energy is:
\be
H= \int d^Dx \(( \sum_{\mu=1}^D \sum_{\alpha=1}^N \left({\partial
\phi_\alpha
   \over \partial x_\mu}\right) ^2
  +{\mu \over 2} \sum_{\alpha=1}^N \phi_\alpha^2
  + \int dx \ V(x,\bphi(x)) \)).
\ee
where  $V$ is a gaussian random potential, the correlations of which
 are taken as:
\be
\overline{V(x,\bphi) V(x',\bphi')}= -N \delta(x-x') f\(( {(\bphi-\bphi')^2
\over N}\)) \ ,
\ee
with:
\be
f(b)={(\theta +b)^{1-\gamma} \over 2(1-\gamma)}.
\ee
We assume a Langevin dynamics:
\be
{\partial \phi_\alpha(x,t) \over \partial t} = - {\partial H \over \partial
\phi_\alpha(x,t)} + \eta_{\alpha}(x,t),
\label{langevin}
\ee
where $\eta$ is a white noise with $<\eta_\alpha (x,t) \eta_{\alpha'} (x',t')>
= 2 T \delta_{\alpha \alpha'} \delta(x-x') \delta(t-t')$. This dynamics
 can be studied
by usual field theoretical techniques \cite{sz} which are reviewed,
 in the present
context, in \cite{kh1}.  We present an alternative derivation
of the equations, based on the cavity method \cite{mpv}, in the appendix.
 For the OED, we find that, in the large N limit,
 the correlation:
\be
C(x,t;x',t')= <{1 \over N} \sum_\alpha \phi_\alpha(x,t) \phi_\alpha(x',t')>
\ee
and the response:
\be
r(x,t;x',t')=<{1 \over N} \sum_\alpha {\partial \phi_\alpha(x,t)
\over \partial \eta_{\alpha} (x',t')}>
\ee
satisfy the following equations: For $t>t'$:
\be
{\partial r(x,t;x',t') \over \partial t}= (\Delta_x-\mu) r(x,t;x',t')
+\int_0^t ds  \ m(t,s;x)(r(x,t;x',t')-r(x,s;x',t'))\ ,
\label{eqr}
\ee
\bea
{\partial C(x,t;x',t') \over \partial t} &=& (\Delta_x-\mu) C(x,t;x',t')
+ 2 \int_0^{t'} ds \  w(t,s;x) \ r(x,t';x',s)
\\
&+& \int_0^t ds \  m(t,s;x) \ (C(x,t;x',t')-C(x,s;x',t'))\ ,
\label{eqc}
\eea
and:
\bea
{1 \over 2}{d C(x,t;x',t) \over  dt}&=& (\Delta_x-\mu) C(x,t;x',t)
+  2 \int_0^t ds  \ w(t,s;x) \ r(x,t';x',s)
 \\
&+& \int_0^t ds \  m(t,s;x) \ (C(x,t;x',t)-C(x,s;x',t))\ +T \ .
\label{eqcdiag}
\eea
In these equations, we have used the following notations:
\bea
w(t,t';x)&=&f'(b(t,t';x)), \ m(t,t',x)= 4 f''(b(t,t';x)) r(x,t;x,t') \\
 b(t,t';x)&=&C(x,t;x,t)+C(x,t';x,t')-2C(x,t;x,t') \ .
\label{def}
\eea
This set of equation is causal. The boundary conditions on $r$
are $r(x,t,x',t^-)=\delta (x-x')$. Given an initial condition $C(x,0,x',0)$,
it has a unique solution.

In the following we shall  concentrate
on the toy model, $D=0$, where the space dependence in
these equations is dropped.
We note that in this limit the model,
described by the simple Hamiltonian $H=(1/2) \mu\sum_\a
\phi_\a^2 + V(\phi_1,...,\phi_N)$, admits another
 interesting interpretation as a spin-glass.
The components $\phi_a$ can be thought as soft spins in
a quadratic well, interacting {\it via} the random potential
 $V$. In particular, in its
 spherical version,  i.e. taking the constraint ${1 \over N }
\sum_\a \phi_\a^2=1$,
it is possible to choose
the values of $\theta$ and $\gamma$  such as to obtain the
spherical $p$-spin model considered in \cite{crihorsom,cuku}.

\section{Static replica solution}
We briefly review here the results of the static r.s.b.
 approach for the toy-model
($D=0$), concentrating  on quantities that  we will study in dynamics.
We keep to the case of "long range" disorder correlation
$\gamma<1$
 where the replica symmetry breaking is of the
full continuous kind.
The  equilibrium statistical mechanics   of the model has been studied in
\cite{mp,mptoy} for the special case $\theta=0$.
 In dynamics a non-zero $\theta$ is
needed to regularize the correlations of the potential at short distance.
The results of  \cite{mp,mptoy} generalize as follows.
At high temperature the system is ergodic and replica symmetric,
and the equilibrium is characterized by the correlations
\beqna
{1\ov N}\sum_{\a}\ol{\la \phi_\a^2 \ra}_{Gibbs}
 =\tilde{q}={T\ov \mu}+{1\ov \mu^2}
(\theta+{2T\ov \mu})^{-\gamma}
\\
{1\ov N}\sum_{\a}\ol{\la \phi_\a \ra^2 }_{Gibbs}=q={1\ov \mu^2}
(\theta+{2T\ov \mu})^{-\gamma}
\label{resrs}
\eeqna
where by angular brackets we have denoted
 the thermal average and by an overline
 the disorder average.
The energy is given by
\be
E={\mu\ov 2}\tq+{1\ov T}[f(0)- f(2(\tq-q))]
\ee
At a  critical temperature $T_c$,
\be
T_c={\mu\ov 2 }\l(-\theta+{\mu^2 \over 2 \gamma}\r)^{({-1 \over 1+\gamma})}
\label{tc}
\ee
there is a phase transition and replica symmetry is broken.
The thermodynamics of the system is fully specified by
$\tilde{q}={1\ov N}\sum_{\a}\ol{\la \phi_a^2 \ra }$ and by a function
$q(u)$, $u\in [0,1]$. Standard arguments  from the  mean field theory
of spin glasses \cite{mpv}, imply breaking of ergodicity and the existence
of many pure states, whose correlations are characterized by
 a non trivial $P(q)$ defined as the overlap
 distribution for two copies of the system with identical
 realization of the random potential $V$:
\be
P(q)= \overline{<{1 \over N} \sum_{\alpha=1}^N \delta(\phi_\alpha
\psi_\alpha -q)>_{Gibbs}}={\d u(q)\ov \d q}
\label{pdeq}
\ee
where $u(q)$ is the inverse function of $q(u)$.
The  order parameter function $q(u)$ is:
\be
q(u)=
\l\{
\begin{array}{ll}
q_0 & u\le u_0\\
\tilde{q}+{\theta\ov 2} -{1\ov 2}\l({\sqrt{2\gamma}\ov 1+\gamma}{u\ov T}\r)^{
2/(\gamma-1)} &u_0\le u\le u_1
\\
q_1 &u_1\le u\le 1
\end{array}
\r.
\label{qdiu}
\ee
where
\beqna
 q_0&=&{1\ov 2\gamma}\l({\mu^2\ov 2\gamma}\r)^{-1/(1+\gamma)}
\\
u_0&=&
T(1+\gamma)\l(\mu\r)^{(1-\gamma)/(1+\gamma)}\l(2\gamma\r)^{-1/(1+\gamma)}
\nonumber
\\
u_1&=&{T\ov \sqrt{2\gamma}}(1+\gamma)(\theta+2(\tilde{q}-q_1))^{(\gamma-1)/2}
\nonumber\\
\tilde{q}&=&{1+\gamma\ov 2\gamma}\l({\mu^2\ov
2\gamma}\r)^{-1/(1+\gamma)}-\theta/2
\eeqna
and $\tilde{q}-q_1$ is the solution of the equation
\be
\tilde{q}-q_1={T\ov \sqrt{2\gamma}}(\theta+2(\tilde{q}-q_1))^{(1+\gamma)/2}.
\label{q1r.s.b.}
\ee
{}From the knowledge of $\tq$ and $q(u)$ all the physical quantities at
equilibrium can be calculated, for example
the energy is:
\be
E={\mu\ov 2}\tq+{1\ov T}[f(0)-\int_0^1 \d u \ f(2(\tq-q(u)))]
\ee
The results presented here for the $q(u)$ have also been obtained in
\cite{kh2} in the ED approach with Sompolinsky ansatz, which,
as we have already
remarked, reproduces the algebra of the  r.s.b. approach.

\section{Analytic study of the asymptotic regime}

The scope of this section is to study the behaviour of the solution of
the dynamical equations
(\ref{eqr},\ref{eqc},\ref{eqcdiag})
in the "asymptotic" limit. This limit is defined as $t=t_w+\tau,t'=t_w$,
with $t_w \to \infty$ while $\tau$ is kept fixed.
For the sake of the simplicity of the presentation, we shall
present the whole analysis in the case $D=0$. The generalization
of the analytic results to higher dimensional problems is straightforward.
We rewrite here, just for graphical transparency, the dynamical
equations (\ref{eqr},\ref{eqc},\ref{eqcdiag}) for $D=0$.
\beqna
{\partial r(t,t') \over \partial t}&=& -\mu r(t,t')
+\int_0^t ds  \ m(t,s)(r(t,t')-r(s,t'))\ ,
\\
{\partial C(t,t') \over \partial t} &=&-\mu C(t,t')
+ 2 \int_0^{t'} ds \  w(t,s) \ r(t',s)\nonumber\\
&+& \int_0^t ds \  m(t,s) \ (C(t,t')-C(s,t'))\ ,
\nonumber\\
{1 \over 2}{d C(t,t) \over  dt}&=& -\mu C(t,t)
+  2 \int_0^t ds  \ w(t,s) \ r(t,s) \nonumber\\
&+& \int_0^t ds \  m(t,s) \ (C(t,t)-C(s,t))\ +T \ ,
\label{em}
\eeqna
with
\bea
w(t,t')&=&f'(b(t,t')), \ m(t,t')= 4 f''(b(t,t')) r(t,t') \\
 b(t,t')&=&C(t,t)+C(t',t')-2C(t,t') \ .
\label{def0}
\eea
For future reference we also give the formula for the energy:
\be
E(t)={\mu\ov 2}C(t,t)-2\int_0^t \d s \ f'(b(t,s))r(t,s)
\label{energy}
\ee

Let us make the reasonable  assumption, supported by the numerical
integration below, of the existence of an asymptotic
regime for $t,t'\to\infty$ keeping $\tau=t-t'$ finite.
Namely we will suppose the existence of  the two limiting functions
\be
r_{as}(\tau)=\lim_{t'\to\infty}r(t'+\tau,t')
\ee
\be
C_{as}(\tau)=\lim_{t'\to\infty}C(t'+\tau,t').
\ee

Taking the limit of the dynamical equations (\ref{em})
in the asymptotic regime, we get the
non causal equations
\bea
{d\bas \over d\tau} &=& (-\mu+M_{as}+\Mb)\bas(\tau)
-\int_0^\tau d\tau ' \
m_{as}(\tau-\tau ') \ \bas(\tau ') \ + 2 T \\ \nonumber
&-& \int_0^\infty d\tau'
\l[ m_{as}(\tau+\tau ')- m_{as}(\tau')\r] b_{as}(\tau ') + 4  \l[
w_{as}(\tau+\tau
')
-w_{as}(\tau')\r] r_{as}(\tau ') , \\
{d\ras \over d\tau} &=& (-\mu+M_{as}+\Mb)\ras(\tau)-\int_0^\tau d\tau ' \
m_{as}(\tau-\tau ') \ \ras(\tau ') \ , \\ \nonumber
C_{as}(0) &=& {1 \over \mu-\Mb} \l(T+{1 \over 2} \int_0^\infty ds \ m_{as}(s)
\ b_{as}(s)+2 \int_0^\infty ds \ w_{as}(s) \ r_{as}(s) \r)
\label{eqas}
\eea
where for convenience we have written the equation for the correlation in
terms of $b_{as}(\tau)=2[C_{as}(0)-C_{as}(\tau)]$
instead of $C_{as}(\tau)$ and we have denoted
\be
M_{as} \equiv \int_0^\infty\d\tau  \ m_{as}(\tau),
\label{mas}
\ee
\be
\overline{M} \equiv \lim_{t\to\infty}\int_0^t\d s \ m(t,s)-M_{as}.
\label{anomdef}
\ee
The functions $m_{as}$ and  $w_{as}$ are defined in a way similar to $m$
and $w$ in (\ref{def0}), but using the asymptotic correlation and response.

The term
$\overline{M}$, which we will call "anomaly" in the following,
 is the term
which couples the asymptotic  time regime  ($\tau=t-t'$ finite)
to the non asymptotic ones. The equations
(\ref{eqas})
are identical to those which appear in the ED studied by Kinzelbach and
Horner \cite{kh2}. The only difference lies in the interpretation of the
anomaly: In ED it is supposed to be due to the response of the system to
some perturbations taking place on infinite time scales. This is not
easy to define, since the regularization of these diverging time
scales by using a finite volume system in principle invalidates the
derivation of the dynamical equations (\ref{eqas}). The definition
(\ref{anomdef}) of the anomaly in OED is very clear.

Let us now briefly quote the following results
from the study of ED in \cite{kh2}: One may search a solution of
the asymptotic equations (\ref{eqas})
which satisfies the fluctuation-dissipation-theorem (f.d.t.):
\be
T r_{as}(\tau)=-{\partial\over \partial \tau}C_{as}(\tau)={1 \over 2}
{\partial\over \partial \tau}b_{as}(\tau).
\ee
Assuming the f.d.t., the asymptotic equations simplify to:
\be
{d\bas \over d\tau}= 2T-\bas(\tau) \l(\mu-\Mb-\int_\tau^\infty ds \ \mas(s)
\r)+\int_0^\tau ds \ \mas(s) (\bas(\tau)-\bas(s))
\ee
The condition for the existence of a  monotonous
solution $b(t)$ to this equation is that
\be
\bas(\infty)={2T \over \mu-\Mb}<b_m \ ,
\ee
where $b_m$ is the point where the function of $b$: $T/b-f'(b)/T$ is minimal.
There are two regimes: at temperatures above the critical temperature
$T_c$ which equals the value (\ref{tc}) computed within the static approach,
there exists a solution when the anomaly $\Mb$ is zero. This solution
agrees with the static replica symmetric results (\ref{resrs}):
\be
\bas(\infty)={2T \over \mu}=2(\tilde{q}-q) \ , \ C_{as}(0)=\tilde{q} \ ,
\ee
At low temperatures, $T<T_c$,  there is no solution
satisfying the f.d.t. relation
 if $\ol{M}=0$. For such a solution to exist one needs a non
zero anomaly: $\Mb \le \mu-2T/b_m <0$. The special choice (named
"postulate of marginal stability" in \cite{kh2}) of the anomaly:
\be
\Mb = \mu-2T/b_m
\label{an}
\ee
leads to an asymptotic correlation
$\bas(\infty)=b_m$, which is equal to the static result: $2(\tilde{q}-q_1)$
computed within the static approach with r.s.b. (\ref{q1r.s.b.}).
Similarly, one gets $C_{as}(0)=\tilde{q}$

We can summarize this discussion about the asymptotic dynamics
in the low temperature phase as follows: In view of the
static analysis, and its interpretation in terms of ergodicity
breaking,
it is reasonable to assume the existence of an asymptotic regime,
obeying the f.d.t., and such that the two following static correlations
are recovered: $C_{as}(0)=\tilde{q}, \ C_{as}(\infty)=\tilde{q}-q_1$.
However for such a regime to exist one needs a non zero value
of the anomaly. In the next section we present some numerical
results which confirm the validity
of these  assumptions, in section 6 we study the implications
of the existence of an anomaly in terms of aging.

\section{Numerical study of the asymptotic regime}

While the set of assumptions which have been put forward at the end
of the previous section look very reasonable, they still deserve
 a confirmation. (In fact some models have been found, such as the spherical
spin glass with p($\ge 3$) spin interactions, where even the values of the
critical temperatures found in the static and dynamic approaches are
different \cite{kirtir,cuku,crihorsom}. It is believed that this effect
is related to the fact that the replica symmetry
breaking is first order in these models.) If these assumptions are correct,
it means that the system of causal first order equations (\ref{eqr},
\ref{eqc})
contains the static solution with full replica symmetry breaking,
which is in itself an interesting observation.

In this section we present a numerical study of
the dynamical equations
(\ref{em}). Our aim is to study the low temperature phase of the model
 comparing the result of the integration with the static  solution and the
asymptotics of
the previous section. The values of the parameters appearing in the
 Hamiltonian
have been chosen equal to $\gamma=1/2$, $\mu=1/8$,
$\theta=5$. With this choice
the critical temperature is $T_c=.658$, and for $T\le T_c$
the static correlations take the value $\tq=21.5$

The discretization of (\ref{em}) was chosen to be the simple one
induced by the discretization of the  Langevin equation (\ref{langevin})
with the Ito
convention.
We have solved the discrete equations with time steps $4 h$, $2 h$,
and $h$, and extrapolated the correlation and response to
$h=0$ by a second degree polynomial.  $h$  was chosen in such a way
that this extrapolation does not differ too much from the linear
extrapolation of the data at $2 h$ and $h$. In this way with $h=.3$ we where
able to reach
times of the order of 1000. We also performed the integration of the equation
for longer times for some particular value of $h$, as we will specify in the
following.
In  most of the simulation the initial condition $C(0,0)=21.5$ was taken.
We have checked that the dynamics in the asymptotic region does not depend
on this choice.

We have integrated the system (\ref{em}) for $T=.5$ and $T=.2$. A run was also
performed at $T=3>T_c$. With this last run
we checked that in the high temperature phase the OED simply corresponds
to the relaxation into the unique equilibrium state described by the r.s.
statics. Coherently we find that $C(t,t)$ tends exponentially to
its r.s. value
$\tq_{rs}=32.8$, and the energy to $E_{rs}= 0.368$.
In the low temperature phase the situation changes. The asymptotic
extrapolation for $C(t,t)$ and $E(t)$ become incompatible
 with the r.s. values.
As a first approximation, the behaviour of the equal
time correlations
$C(t,t)$ is compatible with  a power law
approach to its asymptotic value (with an exponent, deduced
from the behaviour of $dC(t,t)/dt$, equal to $-.73 \pm .05$)
\cite{fm1}.
When one uses this power law fit  in
order to extrapolate $C(t,t)$ to infinite $t$, it yields
the result  $21.4 \pm .1$
which is in agreement with the r.s.b. prediction $21.5$. For lower
temperatures this procedure is less precise, and there are clearly
corrections to the simple power law behaviour of $dC(t,t)/dt$.
Better estimates for the asymptote  are obtained fitting
the time derivatives of
$C(t,t)$ and $E(t)$ with functions depending on three parameters:
\beqna
f_1(t)&=&a_1 t^{-a_2}(1+{a_3\ov t})
\\
f_2(t)&=&a_1 t^{-a_2}(\log(t))^{a_3}(1-a_2+{a_3\ov \log(t)}).
\label{fit}
\eeqna
In the time window we reach, these two fits give comparable errors, but
also comparable estimates for the asymptote (after integration of the fits).
For instance we show in Fig.1a
an estimate of the large time limit , $C_{\infty}$,
of the autocorrelation C(t,t), from the numerical solution of the
dynamical equations with a grid size $h=1.2$.
The derivative $dC(t,t)/dt$ has been fitted to the function
$f_2$. For each time $t$, $C_{\infty}$
is approximated by $C(t,t)$ plus the integral of the fit of the
derivative. The plot gives  $C(t,t)+\int_t^\infty f_2(t') \ dt'$
versus the time $t$. Fig.1b shows that the effect of  the interpolation
at $h=0$ become small at large time. Altogether this procedure
gives $C_{\infty} \simeq 21.49$, with an error, due to the fit, the
extrapolations, which we estimate subjectively to $\pm .05$. This
is quite compatible with
the analytical result from r.s.b., $C_{\infty}=\tq=21.5$.
In Fig.2 we give the analogous plots for the energy. The correction due to
the finite grid size do not vanish at long times and must be
incorporated. We get as a final result:
$E_{\infty}=-1.366 \pm .02$, in very good agreement with the r.s.b.
computation: $E_{\infty}=-1.3660$.
Similar results can be found at a temperature $T=.2$.
Probably the best evidence for the convergence of $C(t,t)$ to $\tq$
is obtained
considering the quantity
\be
A(t)=\tq(1-r(t,0))-C(t,t).
\ee
and observing  that $r(t,0)$, the response at time $t$ to
a change in the field at time zero, should tend to zero at large time.
So if $C(t,t)$ converges to $\tq$, $A(t)$ must go to $0$ at large times.
 In Fig.3 A(t),
as well as $B(t)=\tq-C(t,t)$ are plotted on a log-log scale for T=.2.
A pure power law fit gives:
\beqna
A(\infty)&=&.04
\\
B(\infty)&=&.3,
\eeqna
the quality of this  two parameter fit on $A$ is comparable with the ones
we had on $C$ with
logarithmic or power law corrections.

Let us now turn to the study of the asymptotic functions $b_{as}$ and $r_{as}$.
In Fig. 4a we plot for $T=.5$ the response $r(t_w+\tau,t_w)$ versus $\tau$
for various values of the waiting time $t_w$. We also give the result
$r_{as}(\tau)$ of a 3 parameter power law extrapolation of these
data at $t_w=\infty$. The same is done in Fig.4b for the correlation
$b(t_w+\tau,t_w)=C(t_w+\tau,t_w+\tau)+C(t_w,t_w)-2C(t_w+\tau,t_w)$.
According to the statics, the correlation should go to
$lim_{\tau \to \infty} b_{as}(\tau)=2(\tilde q-q_1)=6.068$. It is possible
to see directly that the data is compatible with this asymptota, with
a power law approach. However, in view of the relatively short times
$\tau$ accessible here (keeping $\tau<<t_w$), we prefer to use a different
approach which is the comparison to an analytic study of the
asymptotic equations.
In Fig.4 the limiting functions obtained from a power law interpolation are
compared
to those obtained from the numerical integration of the asymptotic equations
(\ref{eqas})
with the anomaly set to its "marginal" value (\ref{an}).
 The agreement is very good.
This confirms that the asymptotic dynamics coincides with
the ED on finite timescales, and agrees
 with the static r.s.b.
results.

\section{The non asymptotic regime: aging}

We now turn to the non asymptotic times. From the previous sections we know
that there exists a non zero "anomaly". This means that the decay
of the response $r(t,s)$ at large $t-s$ is slow. More precisely, it
implies that the integrated response at
a large time $t$, $\int_0^{t} r(t,t-\tau) \d\tau$,
receives some finite contributions from  time differences $\tau$ which
 diverge when $t$ goes to infinity. We define such a situation as a
situation of aging. This definition is compatible with the ones used so far.
It basically means that even at large times the physics of the system
depends on its previous history. Besides the usual asymptotic regime
$t \to \infty, \ t' \to \infty, \ t-t' $ finite,  there exist
other "crossover regimes",
in which the limit $t,t'\to \infty$ is taken in a
 different way. The asymptotic regime cannot be decoupled from these
other regimes.

We now propose a solution of the dynamical equations, giving the
correct result for the anomaly, in the
non asymptotic regime.  Basically we propose a reformulation of
the Sompolinsky Ansatz  \cite{sompo,kh2}
in the context of OED. The main difference is that here
we do not impose temporal homogeinity in the equations {\it ab initio}.
The diverging time scales of Sompolinsky's approach,
 needed for the system to cross the diverging barriers,
are here substituted by some function of the  waiting time $t_w$,
which provides a natural cut-off for the theory.
A simple version of this scenario, including one single crossover
domain (corresponding to a single step of r.s.b.), had been found
by Cugliandolo and Kurchan in the spherical p-spin
model \cite{cuku}. Recently they have also proposed a similar scenario
for the OED of the Sherrington
Kirpatrick model  close to its critical temperature \cite{cukusk}.
Let us  perform the limit $t,t'\to \infty$ by
dividing the octant $t'\le t$ into non-overlapping crossover domains.
 A crossover domain
${\cal D}_u$ is defined, using an increasing function $h_u(t)$, as the set of
times $t,t'$ which are both large, but keeping the ratio $h_u(t')/h_u(t)
=exp(-\tau)$ fixed, with $\tau \in ]0,\infty[$.\footnote{ The index $u$ of the
domains
should at first be taken as a discrete variable, in a procedure analogous to
that of statics in which one considers first a finite number of r.s.b.
and then passes to the continuum limit. This is familiar to the reader both
from the static r.s.b. approach and from the ED, and it will be not repeted
here.
We just mention that $u$
will turn out to be a continuous variable in the interval [0,1].}
 Suppose that in the
crossover domain
${\cal D}_u$ one has:
\be
b(t,t')=\hat b_u(\tau) \ , \ r(t,t')={d \ \ln[h_u(t')] \over dt'}
\hat r_u(\tau) \ .
\ee
Then the contribution to  the anomaly
$\int_0^t ds \  m(t,s)$ from all the times $s$
such that $s $ and $t $ are in ${\cal D}_u$ is finite
and equal to
\be
\int_0^\infty\d\tau 4 f''(\hat{b}_u(\tau))\hat{r}_u(\tau)
\ee
which is independent on the function $h_u$.

In a simple problem
like for instance the high temperature phase, there should exist
a single crossover domain, the asymptotic one defined by $h(t)=e^t$. In
a glass phase, we can have a  relatively simple scenario in which there
exists, beside the asymptotic domain, another one defined by some other
function $h(t)$. Such a case (with $h(t)=t$)
 has been found recently \cite{cuku}. But
one can also have some systems with many crossover domains.
 The condition we impose is that they do not overlap. We can index them
by a parameter $u$ such that, if $w<u<v$ and the points ($t,t'$) belong
to ${\cal D}_u$, then $h_v(t')/h_v(t)=0$ and $h_w(t')/h_w(t)=1$. A
possible choice leading to such a behaviour would be for instance
$h_u(t)=exp(t^u)$. With this choice the points ($t,t'$) belong to
${\cal D}_u$ when $t'=t-t^{(1-u)}\tau/u$.

The alert reader will have recognised
in this scenario a hierarchical stucture which is reminiscent of the
ultrametricity assumption underlying both the statics and the equilibrium
dynamics \cite{mpstv}.
 We have here a hierarchy of time crossovers. Considering
three times $t''<t'<t$, one sees that, if ($t,t'$) belongs to the
crossover domain ${\cal D}_u$ and ($t',t''$) belongs to ${\cal D}_v$,
then ($t,t''$) belongs to ${\cal D}_{inf(u,v)}$, which is an
ultrametric inequality, and obviously implies ultrametric
relations for the corresponding correlation functions.

The dynamical equations can be solved within this scenario because
one can forget the time derivatives in the dynamical equations.
The existence of
an asymptotic regime in which $\lim_{\tau\to\infty}{\partial C_{as}(\tau)
/\partial \tau}=0,$  implies
 that in the crossover regimes
${\partial C(t,t')\over \partial t'}\to 0 $ while ${\partial r(t,t')\over
\partial t'}$ tends to zero more rapidly then $r(t,t')$. The l.h.s.
of (\ref{em}) can be neglected in this situation and
 the problem  becomes invariant under the family of transformations
\beqna
C(t,t')&\to&C(h(t),h(t'))\nonumber\\
r(t,t')&\to &{\d h(t')\over \d t'}r(h(t),h(t'))
\eeqna
for any monotonically increasing function of time  $h(t)$.
Any non trivial solution will break this invariance, consequently from a given
solution we can generate a whole "orbit" of equivalent ones
 just reparametrizing the time.
  As we have already remarked,
the solution of (\ref{em}) is unique at any  finite times $t,t'$.
The appearence of this invariance seems somewhat artificial; among
all these possible solutions, only one can be the asymptote of the finite
time dynamics. At this stage it is an open problem what is
the  choice  which will be picked up by the dynamics.

The ambiguity due to the time reparametrization invariance of the asymptotic
equations reflects in the fact that the equations for  $\hat b_u$
and $\hat r_u$ are {\it independent} of the choice
of all the arbitrary
functions $h_u(t)$. In fact these equations are identical to
those derived in ED on diverging time scales;
this set of equations has been shown  \cite{kh2}
to possess solutions satisfying
the "quasi f.d.t." relation:
\be
u {\d \hat{b}_u\over \d\tau} =2 T \hat{r}_u(\tau).
\ee
Denoting  $b^+_u=\hat{b}_u(0)$ and $b^-_u=\hat{b}_u(\infty)$ one has for
adjacent domains indexed by $u<u'$, $b^+_u=b^-_{u'}$. Within the OED,
we find that the dynamical correlations are related to the
static order parameter function $q(u)$ by the formula
\be
b^-_u=2(\tilde{q}-q(u)).
\ee
With these ingredients we reproduce the
algebra of the static replica solution, which gives the value
(\ref{an}) for the
anomaly.
In each domain, apart from the asymptotic one, the variation
with $\tau$ of the functions
$\hat{b}_u(\tau)$ is infinitesimal, and $q(u)$ becomes
the continuous function given by (\ref{qdiu}).

As we stressed before, this solution can be understood as a reinterpretation
of the ED solution, and of the static r.s.b. solution. With respect to the ED
solution, the main advantage is that the diverging
time scales have been replaced
basically by some powers of the waiting time. Unfortunately it does not solve
the second problem of ED, namely the invariance through reparametrizations
of time which implies that one looses all the physical
(crossover) time scales. We stress that this is only a problem of the family
of solutions that we have introduced. This problem is not intrinsic to the
OED itself. On the contrary, in the real OED problem there is a unique solution
to the dynamical equations. This solution might go
asymptotically to one of the solutions we have presented here (choosing
dynamically a set of functions $h_u(t)$), or it might even converge
to some other asymptote. So far we have not been able to answer this problem
analytically. So we shall now propose some
numerical checks which proceed through the
numerical solution of the dynamical equations.

The numerical test of this family of  solutions might seem hopeless
insofar as they depend on an arbitrary set of functions $h_u(t)$ which allow
for a reparametrization of time. We shall call such a set a choice
of gauge. In order to decide whether the asymptotic solution belongs
to our family, we propose to use criteria which are gauge independent.
One possibility is to use some integrated quantities  like the
"dynamical moments" introduced in \cite{cuku}:
\be
\label{mom}
{\cal{C}}_k(t) \equiv k \int_0^t ds \ Tr(t,s) C(t,s)^{k-1},
\ee
Within our scenario of hierarchical crossover domains,
 these moments should have a large time limit
given by:
\be
\lim_{t \to \infty} \cC_k(t)=\tilde{q}^k-\int \d q P(q)q^k.
\label{momstat}
\ee
Another possibility consists in the introduction of the function:
\be
U(t,t')={T r(t,t')\over {\partial C(t,t')\over\partial t'}}
\label{fdr}
\ee

In the crossover regime, where the f.d.t. relation holds, $U$ takes the value
$U(t,t')\equiv~1$ at large times, while, in the crossover
domain it gives us a measure of the
violation of f.d.t.. We shall call this function the fluctuation dissipation
(f.d.) ratio. The gauge invariant prediction of the hierarchical
crossover domains scenario is that, if one plots  the f.d. ratio $U$ as a
function of
the time $t$
along the lines of fixed correlation  $C$, its value at large times is equal to
$u(q)$, the inverse of the order parameter function. Let us make this
statement more precise: we first observe that for fixed
(and large enough) $t$, $C(t,t')$
is a monotonously increasing
function of $t'$. This allows to define the function $t'(q,t)$ as the time
$t'$ such that $C(t,t')=q$. The
prediction is that:
\be
U_d(q) \equiv \lim_{t \to \infty} U(t,t'(q,t)) = u(q)=\int_0^q dq' \ P(q') \ ,
\label{pred}
\ee
which is
the inverse of the order parameter function defined in (\ref{qdiu}).
We have been able to obtain the following general results on the f.d ratio.
It is easy to show that $C(t,0)=r(t,0) C(0,0)$.
Under the reasonable assumption $\lim_{t \to \infty} r(t,0) =0$,
one gets that $\lim_{t \to \infty} C(t,0) =0$, and it is easy to deduce
that
$\lim_{t \to \infty} U(t,t'=0) =0$. We have seen numerically, but we have not
been able to prove, that for large enough $t$, $U(t,t')$ is an
increasing function of $t'$. Together with the f.d.t. result
in the asymptotic regime $\lim_{t \to \infty} U(t,t) =1$, this shows
that $U$ tends to a probability at large times.

We have tried to use the simple dicretization algorithm described in the
previous section to study these aging effects (with the same values
of $h$). Although we shall see that
the times we have reached do not
allow  to draw definitive conclusion on the crossover regimes, we think it
is worth to present some of the data, in order to see what happens
on relatively
short times, and to give an idea of the type of computing effort
 which will be needed in order to solve this problem.
 The values of the
parameters are $\gamma=.5,\theta=5., \mu=.125,T=.5, C(0,0)=0.$
We have checked that the errors due to the discretisation and interpolation
procedures are negligible on the scales of the figures.

We first present some confirmation of the existence of the
aging effect. In Fig.5 we plot the
"thermoremanent magnetization" which
we define as:
\be
\mag(\tau,t_w)=\int_O^{t_w} ds \ r(t_w+\tau,s)
\label{magrem}
\ee
The plot shows  $\mag(\tau,t_w)$ versus $\tau$ for  fixed values
 of the waiting
time $t_w$, on logarithmic scales.
On these time scales, one clearly sees an aging effect which
is qualitatively similar to the one observed in experiments
\cite{AGESWE,AGEUSA,AGEFRA,AGETWO} and numerical simulations
\cite{anderson,rieger,CUKURI}
in spin glasses. The effect is confirmed in Fig. 6 which plots
the normalised correlation
$C(t_w+\tau,t_w)/C(t_w,t_w)$ versus $\tau$, at fixed $t_w$. We
have observed that the curves  do not scale very well as functions of
$\tau/t_w$.

We have tried to test the hierarchical solution by some studies
 of gauge invariant
quantities. We first study the dynamical moments (\ref{mom}).
 The first moment
$\cC_1$ satisfies a  kind of Ward identity (related to the translational
invariance of the distribution of the random potential):
\be
r(t,0)=1-{\mu \over T} \cC_1(t) \ .
\label{ward}
\ee
(a simple proof consists in showing that the two sides
of this equality satisfy the same first order differential equation
in time, with the same initial condition).
As  $r(t,0)$  should vanish at large times, this implies
that $\lim_{t \to \infty} \cC_1(t) = T/\mu =\tilde{q} -\int dq P(q) q$.
Numerically we have checked  that the Ward identity (\ref{ward})
is satisfied with a precision of $10^{-5}$, and that the behaviour
of $r(t,0)$ is consistent with a decay to zero. We have computed
numerically the first five moments  $\cC_k(t),k=1,...,5$. In Fig.7
we plot the third moment versus time. Within the hierarchical scenario
one would expect that its large time limit should be given by the
third moment of the static $P(q)$ as in (\ref{momstat}), which
is equal to $9011$ in our case. The inset of Fig.7 shows that the
relative difference of  $\cC_3(t)$ with this value decays approximately
as a power law. However at $t\simeq 800$ the relative difference is still
of order 10 per cent. Fig.8 shows the fifth moment and its approach
to the static value $5.76 \ 10^6$. We consider this data as compatible
with the hierarchical scenario but not really conclusive. As explained
above, a more
detailed analysis of the data consists in studying
 the f.d. ratio (\ref{fdr}) and to test the prediction (\ref{pred}).
In Fig. 9 we plot $C(t,t')/C(t,t)$ versus the time $t$, along lines in
the $t',t$ plane such that $U(t,t')$ is constant, equal to $u_0$. According
to the hierarchical scenario, this quantity should go
at large $t$ to $q(u_0)/\tilde{q}$ defined in (\ref{qdiu}). On this time
scale, we do not see evidence for such a convergence. To summarize,
we consider the results on the moments as encouraging, but the detailed
analysis on the f.d. ratio shows that simulations on much longer time scales
are needed in order to decide on the correctness of the hierarchical solution.

\section*{Conclusions}

In this paper we have studied the off equilibrium dynamics of a disordered
model which represents  on one hand a limiting case of a manifold in a
random environment, on the other hand  a  spin glass with long range
 interactions. The choice of this model  has several motivations. Its
static solution at low temperatures implies a full continuous r.s.b.,
as  for instance in the SK model; this r.s.b. solution is
known in all details. On the other hand, we can write a closed
set of coupled dynamical equations between the correlation and response.
Because of this, we have been able to generalize the analytic solution
of \cite{cuku}
in the aging regime  of the spherical p-spin model
to a full r.s.b. case and to compare to
a numerical integration of the equations. Simultaneously to our work,
Cugiandolo and Kurchan have also extended their analytic solution
to the SK model close to $T_c$ \cite{cukusk}.

 Our analysis is consistent with the existence of
two regimes at large times in the low temperature phase:
an asymptotic regime where time homogeneity and fluctuation dissipation
relations hold,
 and an aging regime where both these properties
are violated.
 These regimes are
similar to the ones observed in experiments and simulations.
 We have found convincing numerical evidence that the
asymptotic regime agrees with the static r.s.b. results and with the
ED results. The  correlations
are those characteristic of a system reaching equilibrium inside
one single  valley. We have shown that these facts imply the
existence of a non trivial aging regime.

We have proposed a family of solutions of the dynamics at large time
in this aging regime, based on a hierarchy of crossover domains. This
solutions solve the problem of the diverging time scales which had to be
introduced in ED. On the other hand several problems are left open.
We have not been able to show that the dynamics converges to one of these
solutions, and {\it a fortiori} we do not know which of them is
picked up. This choice might well depend on the choice of
the Langevin dynamics and of the type of
initial conditions which are used. We have found that the f.d. ratio
tends to a probability law at large times. Longer simulations
are needed to decide whether this probability
law is identical to the static $u(q)$, as implied by the hierarchical
scenario. At the present stage, we believe that it is crucial to carry
out this numerical study. The physical interpretation of the
dynamical probability is also a very important open question.

It would  be interesting to generalize this approach to
systems driven by an external force (charge density waves, vortex
lattices,...),
and to study more subtle effects like those of temperature cycling.
We would also like to point out that this route of OED seems to be a promising
one towards a rigorous study of spin glasses. One should first obtain
a rigorous derivation of the dynamical equations, and then understand the
large time behaviour of these equations. This is certainly not easy, but
it is a well defined mathematical problem and our work suggests
that these coupled dynamical equations contain
in some sense the full r.s.b. solution. A first step towards a rigorous
derivation of the dynamical equations has been taken recently for the SK model
\cite{benarous}

\section*{ Acknowledgments}

We would like to thank J.P. Bouchaud, L. Cugliandolo,
J. Hammann, J. Hertz, H. Kinzelbach,
J. Kurchan, H. Horner, G.Parisi, E. Vincent and M.A. Virasoro
for many interesting discussions.

\section*{Appendix A}
In this appendix we sketch the derivation of the mean field dynamical equations
(\ref{em}) for the toy-model by  the cavity
method \cite{mpv}. This method provides the same results as the functional
derivation
of \cite{kh1}. We include a brief description here because it is maybe more
explicit
on the physical content of the derivation.
Apart from unessential complications, the derivation
could be done similarly for
 the more general equations (\ref{eqr},\ref{eqc},\ref{eqcdiag})
for finite $D$.
The method  involves an induction
over the number dimensions $N$ of the space in which the particle lives,
together with a large $N$ limit.
 We pass from a $N$ dimensional system described by
$\bphi=\{ \phi_1,...,\phi_N\}$ to a $N+1$ dimensional one described by $\bphi$,
plus a new component $\phi_0$.
 In the derivation we  follow a procedure analogous to that which
has been used e.g. to study the statics and the equilibrium dynamics of the
SK model.
We will make crucial use of two hypotheses that {\it mutatis mutandis}
habe been  put
forward in that case. Namely, the applicability of the linear response theory
fort the Langevin equation (LRT), and the fact that the responses
$\de \phi_\a(t)/\de \eta_\b(s) $ can be considered
 small (in a suitable sense) for
$\a\ne\b$. A justification of these  in the case of equilibrium dynamics
is given in \cite{mpv}. For OED, we just assume these two facts. It will be
interesting
to see if similar assumptions are contained in the functional approach, or
whether
these facts can be derived.

Consider the Langevin equation for the toy-model:
\beqna
{\d \phi_\a(t)\ov \d t}=-{\partial H(\bphi(t))\ov \partial \phi_a}+\eta_\a(t)
\nn
\\
\la \eta_\a(t) \eta_\b(s) \ra =2 T\de_{\a\b}\de (t-s).
\label{lang}
\eeqna
If an infinitesimal
perturbation $\de H(\bphi)$ is added to $H$, the perturbed process $\bphi^*(t)$
can be expressed in terms of the unperturbed one $\bphi(t)$ by the linear
response relation:
\be
\phi_\a^*(t)=\phi_\a(t)-\sum_\b\int_0^t \d s \ {\partial  \de H(\bphi(s))\ov
\partial \phi_\b}{\de \phi_\a(t) \ov \de \eta_\b(s)}.
\label{lrr}
\ee

Let us now introduce the new component, and denote by $V_N(\bphi)$ and
$V_{N+1}(\phi_0,\bphi)$ the random potentials for the $N$ and $N+1$ components
systems respectively.
In making this step, the Hamiltonian $H=\mu\bphi^2/2+V_N(\bphi)$ will
undergo the variation
\beqna
\de H(\phi_0,\bphi)=\mu\phi_0^2/2+\de V(\phi_0,\bphi)
\nn\\
\de V(\phi_0,\bphi)=V_{N+1}(\phi_0,\bphi)-V_N(\bphi).
\eeqna
The $\phi^*_\a$ and $\phi_\a$ in (\ref{lrr}) have to be identified
with the $\a$-th component of the position of the particle respectively in
presence and in absence of $\phi_0$.

To study the statistical properties of $\de V$ we can expand in series
the correlations of the potential of the $N+1$ components system
\be
\ol{V_{N+1}(\phi_0,\bphi)V_{N+1}(\psi_0,\bpsi)}=-(N+1)f\(( {[(\bphi-\bpsi)^2+
(\phi_0-\psi_0)^2] \over (N+1)}\)),
\ee
and retain only the terms of the series which do not tend to zero when
$N\to\infty$.

In this way we find:
\beqna
\ol{\de V(\phi_0,\bphi)\de V(\psi_0,\bpsi)}=&-&\l\{f((\bphi-\bpsi)^2/N)-
{(\bphi-\bpsi)^2\ov N}f'((\bphi-\bpsi)^2/N)
\r.
\nn\\
&+&
\l.
(\phi_0-\psi_0)^2 f'((\bphi-\bpsi)^2/N)
\r\}
\eeqna
These formulas can be obtained expanding formally $V_{N+1}(\phi_0,\bphi)$
in powers of $\phi_0$ up to the second order. In this way, denoting
 $b=(\bphi-\bpsi)^2/N$, one  easily
shows that $\de V(\phi_0,\bphi)$ can be written as
\be
\de V(\phi_0,\bphi)=A(\bphi)+B(\bphi)\phi_0+D(\bphi)\phi_0^2
\ee
where $A$, $B$ and $D$ are gaussian random functions with zero averages and
correlations:
\beqna
\ol{A(\bphi)A(\bpsi)}&=&-[f(b)-b f'(b)]
\nn\\
\ol{A(\bphi)B(\bpsi)}&=&O(1/N)
\nn\\
\ol{A(\bphi)D(\bpsi)}&=& -f'(b)
\nn\\
\ol{B(\bphi)B(\bpsi)}&=&2f'(b)
\nn\\
\ol{B(\bphi)D(\bpsi)}&=&O(1/N)
\nn\\
\ol{D(\bphi)D(\bpsi)}&=&O(1/N)
\label{corrnoise}
\eeqna
We can now write the Langevin equation for the zeroth component $\phi_0$
\be
{\d \phi_0(t)\ov \d t}=-\mu \phi_0-B(\bphi^*(t))-2D(\bphi^*(t))\phi_0(t)
+\eta_0(t).
\label{l1}
\ee
Using the LRT we find
\begin{small}
\beqna
B(\bphi^*(t))&=B(\bphi(t))-\int_0^t ds \ \sum_{\a\b}{\partial B(\bphi(t))\ov
\partial \phi_\a}
{\de \phi_\a(t)\ov \de \eta_\b(s)} \nn
\\
&{\partial \ov \partial
\phi_\b}\l[A(\bphi(s))+B(\bphi(s))\phi_0(s)+D(\bphi(s))\phi_0^2(s)\r]
\nn\\
D(\bphi^*(t))&=D(\bphi(t)) -\int_0^t ds \ \sum_{\a\b}{\partial D(\bphi(t))\ov
\partial \phi_\a}
{\de \phi_\a(t)\ov \de \eta_\b(s)} \nn \\
&{\partial \ov \partial \phi_\b}\l[A(\bphi(s))+B(\bphi(s))\phi_0(s)
+D(\bphi(s))\phi_0^2(s)\r].
\label{BD}
\eeqna
\end{small}
At this point we use the hypothesis that  ${\de \phi_\a(t)/\de \eta_\b(s)}$
is  small for $\a\ne \b$. More precisely we  suppose that
 as in the SK mode $1/N^2\sum_{\a\b}{\de \phi_\a(t)/ \de \eta_\b(s)}$
and analogous sums will tend to zero in the large $N$ limit.
One  deduces
 that (\ref{BD}) reads in this limit:

\beqna
B(\bphi^*(t))&=&B(\bphi(t))-\int_0^t\d s \ 4f''(b(t,s))r(t,s)\phi_0(s)
\nn\\
D(\bphi^*(t))&=&D(\bphi(t))-\int_0^t\d s \ 2f''(b(t,s))r(t,s).
\label{pippo}
\eeqna
where $b(t,s)=(\bphi(t)-\bphi(s))^2/N$ and $r(t,s)=(1/N)\sum_\a (\de \phi_a(t)
/\de \eta_a(s))$.

Denoting $m(t,s)=4f''(b(t,s))r(t,s)$ and making use of (\ref{pippo})
the Langevin equation (\ref{l1}) is rewritten  as
\be
{\d \phi_0(t)\ov \d t}=-\mu \phi_0(t)-B(\bphi(t))-2D(\bphi(t))\phi_0(t)
+\int_0^t\d s \ m(t,s)[\phi_0(t)-\phi_0(s)]
+\eta_0(t).
\label{l2}
\ee
(In deriving (\ref{l2}) we have dropped a term
proportional to $D(\bphi(t))\phi_0(t)$ which is negligible because of the
vanishing correlations (\ref{corrnoise}) of $D$  at large $N$.)
 The term
$B(\bphi(t))$ is a random field with zero mean and correlations
$\ol{B(\bphi(t))B(\bphi(t))}=2f'(b(t,s))=w(t,s)$. Therefore equation
 (\ref{l2}) is the usual Langevin equation on one single component, with
a condition of selfconsistence, from which the dynamical equations are
easily derived. We just notice that this derivation shows a property of self
averageness
of the response, namely the fact that
the response function of each component is identical:
$ r(t,s)=<{\de \phi_0(t)/ \de\eta_0(s)}>$

\newpage

\section*{Figure Captions}
\begin{description}
\item{\bf Fig. 1:}
\noindent
a)
An estimate of the large time limit , $C_{\infty}$,
of the autocorrelation C(t,t), from the numerical solution of the
dynamical equations with a grid size $h=1.2$.
The derivative $dC(t,t)/dt$ has been fitted to a power law with
logarithmic corrections (see text). For each time $t$, $C_{\infty}$
is approximated by $C(t,t)$ plus the integral of the fit of the
derivative. The plot gives this estimation, versus the time $t$.
The analytical result from r.s.b., $C_{\infty}=21.5$, is compatible
with the result, when one takes into account the effects due to the
finite value of $h$ (see Fig. 1b) and to the uncertainties of the fit.
\noindent
b)
The difference between $C(t,t)$ computed with a grid size $h=1.2$
and that computed with $h=.8$, plotted versus time in  a Log-Log plot.
This difference seems to extrapolate to zero at large times (with a
power law behaviour).

\item{\bf Fig. 2:}
\noindent
a)
An estimate of the large time limit , $E_{\infty}$,
of the energy E(t), from the numerical solution of the
dynamical equations with a grid size $h=1.2$.
The procedure is the same as that followed for the estimate
of $C(t,t)$ in Fig.1. The analytic result from r.s.b.,
$E_{\infty}=-1.3660$, is compatible with this data when one takes into account
the effect of the extrapolation to $h=0$ (see Fig. 2b, and the text).
\noindent
b)The difference between $E(t)$ computed with a grid size $h=1.2$
and that computed with $h=.8$, plotted versus time.
This difference is well approximated by a power law fit with
an asymptote equal to .0116.

\item{\bf Fig. 3:}
\noindent
The quantities $A(t)$ (continuous line) and $B(t)$ (dotted line)
defined in the text in a Log-Log scale. $A(t)$ is better approximated
by a power law then $B(t)$. A pure power law fit on the last 300 points
over a total of 890  gives $A(t)=.04 +14.8 \  t^{-.57}$ with a relative error
on the whole interval of the order
$\Delta A/A \sim 10^{-6}$ and $B(t)=.29 + 35.1  \ t^{-.57}$
 with $\Delta B/B\sim
10^{-5}$.

\item{\bf Fig. 4:}
\noindent
In a), the response $r(t_w+\tau,t_w)$ versus $\tau$. From top to
bottom, $t_w=432, 504, 576, 648$. Also shown (bottom curve) is the
power law extrapolation of these curves to $t_w \to \infty$, together
with the prediction for $r_{as}(\tau)$ from the asymptotic dynamics (these last
two curves are nearly undistinguishable).
In b), similar curves for the corrrelation $b(t_w+\tau,t_w)$. From top to
bottom,
 $t_w=432, 504, 576, 648$, the extrapolation and the expected result from
the  asymptotic dynamics.

\item{\bf Fig. 5:}
\noindent
The  thermoremanent magnetization $\mag(\tau,t_w)$ defined in
 (\ref{magrem}) versus
$\tau$ for  fixed values of the waiting
time $t_w$, on logarithmic scales. From bottom to top,
$t_w= $38.4, 76.8, 153.6, 307.2, 614.4.

\item{\bf Fig. 6:}
\noindent
The normalised correlation $C(t_w+\tau,t_w)/C(t_w,t_w)$ versus $\tau$,
for  fixed values of the waiting
time $t_w$. From bottom to top,
$t_w= $38.4, 76.8, 153.6, 307.2, 614.4.

\item{\bf Fig. 7:}
\noindent
The third dynamical moment
${\cal{C}}_3(t)$, defined in
(\ref{mom}) versus $t$. The inset is a log-log plot of the relative
 difference between the moment at time $t$ and the prediction from the
 scenario of hierarchical crossovers
 concerning its large time behaviour:
$(9011-{\cal{C}}_3(t))/9011$ versus $t$.

\item{\bf Fig. 8:}
\noindent
The same plot as in Fig. 7, for the fifth dynamical moment versus time,
and its convergence to the theoretical result $5.76  \ 10^6$.

\item{\bf Fig. 9:}
\noindent
The function  $C(t,t')/C(t,t)$ versus  the time $t$, along lines
in the $t,t'$ plane such that the f.d. ratio $U(t,t')$ is constant,
equal to $u_0$. From bottom to top, $u_0=.1,.2,...,.9$.
If the scenario of hierarchical crossovers would hold,
at infinite $t$
the curves with $u0 < .375 $ should extrapolate to $ .744$, the ones with
$u_0 > .411$ should extrapolate to $.859$. There is no such indication
on this time scale.
\end{description}


\begin{thebibliography}{99}

\bi{mpv}
M.~M\'ezard, G.~Parisi, and M.A.~Virasoro, "Spin glass theory and beyond",
World Scientific (Singapore 1987);

\bi{binyou}
K.~Binder  and A.P.~Young, Rev.Mod.Phys. {\bf 58} (1986) 801;

\bi{fishertz}
K.H.~Fischer, J.A.~Hertz, "Spin Glasses" Cambridge University Press, (1991);

\bibitem{AGESWE}
L.~Lundgren, P.~Svedlindh, P.~Nordblad and O.~Beckman,
Phys. Rev. Lett. {\bf 51} (1983) 911;
P.~Nordblad, P.~Svedlindh, L.~Lundgren, and L.~Sandlund,
Phys. Rev. {\bf B33} (1986) 645;

\bibitem{AGEUSA}
R.~V.~Chamberlin, G.~Mozurkevich and R.~Orbach,
Phys. Rev. Lett. {\bf 52} (1984) 867;
R.~Hoogerbets, Wei-Li Luo and R.~Orbach,
Phys. Rev. {\bf B34} (1986) 1719;

\bibitem{AGEFRA}
M.~Alba, J.~Hamman, M.~Ocio and Ph.~Refrigier,
J. Appl. Phys. {\bf 61} (1987) 3683;
F.~Lefloch, J.~Hamman, M.~Ocio and E.~Vincent,
Europhys. Lett. {\bf 18} (1992) 647;

\bibitem{AGETWO}
M.~Lederman, R.~Orbach, J.~M.~Hamman, M.~Ocio and E.~Vincent
Phys. Rev. {\bf B44} (1991) 7403;
J.~M.~Hamman, M.~Lederman, M.~Ocio, R.~Orbach and E.~Vincent
Physica {\bf A} (1992) 278;

\bibitem{kophil} G.J.M. Koper and H.J. Hilhorst, {J. Phys. France}
{\bf 49} (1988) 429;

\bibitem{fishus} D.S. Fisher and D.A. Huse, {Phys. Rev.} {\bf B38}
(1988) 373;

\bibitem{bouchaud}
J.~P.~Bouchaud,
J. Phys. I {\bf 2} (1992) 1705;
J.~P.~Bouchaud, E.~Vincent and J.~Hamman,
J. Phys. France {\bf 4} (1994) 139;

\bi{dedom}
C.~DeDominicis, Phys.Rev. {\bf B18} (1978) 4913;

\bi{sz}
H.~Sompolinsky and A.~Zippelius, Phys.Rev.Lett. {\bf 47} (1981) 359;
  Phys.Rev. {\bf B25} (1982) 6860;

\bi{sompo}
H.~Sompolinsky, \pub{Phys.Rev.Lett.} { 47}{359}{1981};

\bibitem{horner}
H.~Horner, Z.Phys. {\bf B66} (1987) 175; M.~Freixa-Pascual and
H.~Horner, Z.Phys. {\bf B80} (1990) 95;

\bibitem{dfi}
 L.B.~Ioffe,{Phys. Rev.} {\bf B38}
(1988) 5181 ;
 V.S.~Dotsenko, M.V.~Feigelman and L.B.~Ioffe,
Soviet scientific reviews {\bf 15}  (1990) 1;

\bibitem{opper}
H.~Eissfeller and M.~Opper, Phys.Rev.Lett.  {\bf 68} (1992) 2094;

\bibitem{cuku}
L.F.~Cugliandolo and J.~Kurchan, Phys.Rev.Lett.{\bf 71} (1993) 173;

\bibitem{rieger}
H.~Rieger,
J. Phys.  {\bf A 26} (1993) L615;

\bibitem{CUKURI}
L. Cugliandolo, J. Kurchan and F. Ritort,
{\em Evidence of Aging in Spin Glass Mean-Field Models,}
cond-mat/9307001 preprint (July 1993);

\bi{mapa}
E.~Marinari and G.~Parisi,  J. Phys. {\bf A 26} (1993) L1149;

\bi{manifrev}
T. Nattermann and P. Rujan, Int. J. Mod.
 Phys. B3 (1989) 1597; T. Natterman and J. Villain, Phase transitions
11 (1988) 5;

\bi{fglv}
M.V.~Feigelman, V.B.~Geshkenbein, A.I.~Larkin and V.M.~Vinokur,
Phys.Rev.Lett. {\bf 63} (1989) 2303;

\bi{bmy}
 J.P.Bouchaud, M.M\'ezard and J.S.Yedidia,
Phys.Rev.Lett. 67(1991)3840 and Phys.Rev. B46(1992)14686;

\bibitem{kpz}
 M.Kardar, G.Parisi and Y.C.Zhang, Phys.Rev.Lett. {\bf 56}(1986)889;

\bibitem{burgers}
D. Fisher, M.P.A. Fisher, D. Huse, Phys. Rev. B  43, 130 (1991);

\bi{mp}
M. M\'ezard and G. Parisi, \pub{\jpa}{23}{L1229}{1990};
\pub{J. Physique. I}{1}{809}{1991};

\bibitem{villain}
 J.~Villain,B.~Semeria, F.~Lanon and L.~Billard,
\pub{J.Phys.}{C16}{2588}{1983};
J.~Villain \pub{\jpa}{21}{L1099}{1988};

\bi{engel1}
A.~Engel, \pub{J. Physique Lett.}{46}{L409}{1985};

\bi{mptoy}
 M. M\'ezard and G. Parisi, \pub{J.Phys. I France} {2}{2231}
{1992};

\bi{engel2}
 A. Engel,  {\it Replica symmetry breaking in one dimension},
 Gottingen University preprint, 1993;

\bi{fm1}
S.~Franz, M.~M\'ezard, {\it Off-equilibrium glassy dynamics: a simple case.}
LPTENS preprint 93/39, to be published in
{Europhysics Letters}

\bi{kh1}
H.~Kinzelbach and H.~Horner, \pub{J.Phys. I France}{3}{1329}{1993} ;

\bi{kh2}
H.~Kinzelbach and H.~Horner, \pub{J.Phys. I France} {3}{ 1901}{1993};

\bi{cufrakumez}
 L.~Cugliandolo, S.~Franz, J.~Kurchan and M.~M\'ezard,
in preparation;

\bi{crihorsom}
A.~Crisanti, H.~Horner and H-J~Sommers, Z.Phys. {\bf B92} (1993) 257;

\bi{kirtir}
T.R.~Kirkpatrick and D.~Thirumalai, Phys.Rev.Lett. {\bf 58} (1987) 2091,
Phys.Rev. {\bf B38} (1987) 5388;

\bi{cukusk}
L.F.~Cugliandolo and J.~Kurchan, {\it On the out of equilibrium relaxation
of the Sherrington Kirkpatrick model.}
Rome University preprint 977;

\bi{mpstv}
M.~M\'ezard, G.~Parisi, N.~Sourlas, G.~Toulouse and M.A.~Virasoro,
Phys.Rev.Lett. {\bf 52} (1984) 1156; J. Physique {\bf 45} (1984) 843;

\bibitem{anderson}
J.O.~Anderson, J.~Mattson, and P.~Svedlinh, Phys.Rev. {\bf 46} (1992) 8297;

\bi{benarous}
G.~BenArous and A.~Guionnet, {\it Large deviations for Langevin spin glass
dynamics}, preprint Universit\'e de Paris Sud (math\'ematiques) 93.65;
A.~Guionnet, {\it Large deviations for Langevin spin glass
dynamics, part II}, preprint Universit\'e de Paris Sud (math\'ematiques) 93.79.


\end{thebibliography}
\end{document}